\documentclass{iau}
\usepackage{graphicx} 
\usepackage{url} 

\title[IAUS291.~~Pulsars are cool.] 
{Pulsars are cool.  Seriously.} 

\author[S.~Ransom]  
{Scott M.~Ransom}

\affiliation{NRAO, 520 Edgemont Road, Charlottesville, VA 22903, USA
  \\ email: {\tt sransom@nrao.edu}}

\pubyear{2012}
\volume{291}  
\jname{\mbox{Neutron Stars and Pulsars: Challenges and Opportunities after 80 years}}
\editors{J. van Leeuwen, ed.} 

\begin{document}

\maketitle

\begin{abstract}
  Ever since the first pulsar was discovered by Bell and Hewish over
  40 years ago, we've known that not only are pulsars fascinating and
  truly exotic objects, but that we can use them as powerful tools for
  basic physics and astrophysics as well.  Taylor and Hulse hammered
  these views home with their discovery and timing of the spectacular
  ``binary pulsar'' in the 1970s and 1980s.  In the last two decades a
  host of surprises and a promise of phenomenal scientific riches in
  the future has come from the millisecond pulsars.  As our
  instrumentation has become more sensitive and better suited to
  measuring the pulses from these objects, they've given us new tests
  of general relativity, fantastic probes of the interstellar medium,
  constraints on the physics of ultra-dense matter, new windows into
  binary and stellar evolution, and the promise of a direct detection
  of gravitational waves.  These things really are cool, and there is
  much more we will do with them in the future.  
\end{abstract}



Pulsars really are cool.  Not in the temperature sense, given that
their surfaces are at about a million kelvin, but in the other.  These
are city-sized (10$-$12\,km radii) neutron stars with up to twice the
mass of our Sun.  Their central densities are several times higher
than atomic nuclei, so high that our current nuclear and particle
physics cannot accurately predict what goes on deep within these
stars.  They have surface gravities 100 billion times stronger than
the Earth's, making nearby space-time highly curved.  Their magnetic
fields range from 100 million times to a {\em quadrillion} times
stronger than the Earth's --- fields so strong that quantum effects
become important (see Nanda Rea's contribution to these proceedings).
They can spin over 700 times per second, which is faster than racing
car engines rotate and kitchen blenders spin.  They emit
electromagnetic radiation via detailed processes we don't understand
after over 40\,years of hard work and at luminosities, coming only at
the expense of rotation(!), of up to 10,000 times more than the total
output of the Sun.  There is no denying that these are exotic objects.

Yet even if we ignore their exoticness, they are still cool.  The
stories of the original pulsar discovery by Jocelyn Bell and Tony
Hewish \cite[(Hewish et al.~1968)]{hewish68} and of the first binary
pulsar by Joe Taylor and Russell Hulse \cite[(Hulse \& Taylor
1975)]{ht75}, and the ensuing Nobel glories and gaffes, are the stuff
of astronomical legend.  For me though, what really makes pulsars so
cool is how they can be used as tools for a wide variety of physics
and astrophysics problems by the miracle of pulsar timing.  The
``stars'' of these measurements are most certainly the millisecond
pulsars (MSPs).

\section{Millisecond pulsars}

Millisecond pulsars are distinct from the $\sim$2000 ``normal''
pulsars known in that they have been ``recycled'' \cite[(Alpar et
al.~1982 and Radhakrishnan \& Srinivasan\ 1982)]{Alpar82,rs82}.  A
pulsar is born in a supernova and then radiates and spins-down as a
normal (i.e. $\sim$1\,second spin period) pulsar for 10-100\,Myr.  If
the system was in a binary which survives the supernova though, the
secondary star will evolve on Gyr timescales and, when it begins to
ascend the giant branch, transfers mass and angular momentum onto the
long-dead pulsar.  During this period the system is observable as an
X-ray binary.  When the stellar and binary evolution finishes, an MSP
emerges, spinning hundreds of times per second in a nearly perfectly
circular orbit around a white dwarf.  Since the neutron star's
magnetic field is somehow buried by the accretion, from
$10^{12}$\,Gauss down to $10^8$\,Gauss, the new MSP spins down much
more slowly, providing us with a nearly perfect clock visible for
billions of years.

\section{Pulsar timing}

At 8:40AM CST, on August 23, 2012 (which was when I was at this point
in my talk at the IAU in Beijing), the spin period of one of the best
timed MSPs, J0437$-$4715, was exactly 5.7574518556687\,ms with an
error of $\pm$1 in the last digit.  Since the pulsar loses rotational
energy due to its emission of a relativistic wind and electromagnetic
radiation, that last digit increases odometer-like by one every half
hour.  That means that the first 6 digits will remain constant for
about the next millennium!  This stability, and our ability to measure
it via pulsar timing, is why pulsars enable truly revolutionary
measurements.

Pulsar timing is really quite simple in concept.  We unambiguously
account for each and every rotation of a pulsar over a time span of
years.  In practice, we can actually track small fractions of each
rotation, thereby making pulsar timing a precise form of phase
measurement.  With the start of an observation referenced via GPS to
worldwide atomic time standards and time during each observation
tracked using hydrogen masers at the observatories, we can measure the
average times of arrival (aka TOAs) of MSP pulses to better than
1\,$\mu$s, which corresponds to about $10^{-4}$ in rotational phase
$\phi$.  Since an error in frequency is simply $\Delta\phi / \Delta
T$, if we can make measurements like this over time spans $\Delta T$ of
3\,years (i.e. $10^8$\,seconds), our measurement error in the spin
frequency of a pulsar is $10^{-12}$\,Hz, corresponding to about 14
significant figures for MSPs.

We establish timing ``solutions'' for pulsars via a bootstrapping
series of observations scheduled such that we never lose count of the
number of rotations our target pulsar makes.  A dense set of
observations involving several over one or two days, and then several
more spaced over the next week allow us to solve binary orbital
parameters and determine an increasingly precise spin frequency.  As
more time is added to the timing solution, the Earth's orbital motion
allows us to determine highly-precise (down to tens of
micro-arc-seconds for MSPs) astrometric positions and eventually even
proper motions of the pulsars, and the spin-down of the pulsar appears
as an increasing quadratic delay in the pulse arrival times. After a
year, the timing solution is complete, and it includes a precise
position, spin-frequency, and spin-down rate, and many significant
figures in the five Keplerian orbital parameters if the pulsar is in a
binary (orbital period, projected semi-major axis, eccentricity, time
of periastron passage, and the argument of periastron).  Often
characterized by the root-mean-squared (RMS) deviation of the timing
residuals (i.e. TOAs minus model predictions), state-of-the-art timing
solutions are below 100\,{\em nano}-seconds RMS over timescales of 5
or more years.

Published MSP timing solutions are some of the most precise
measurements in all of astrophysics and have enabled science which
would be effectively impossible using other techniques.  A great
example is the fact that MSP timing provides pulsar radial velocity
measurements at ridiculous precisions of a few {\em mm}/s compared to
$\sim$1\,m/s for optical radial velocity planet surveys.  That
precision allowed Wolszczan \& Frail~(1992)\cite[]{wf92} to uncover
the first extrasolar planets around MSP B1257$+$12.  Those planets are
all among the lowest mass exoplanets yet detected, and planet ``A'' is
only twice the mass of the Moon!

\section{A millisecond pulsar renaissance}

Over the past decade, and especially over the past 5 years, the pulsar
field has been focused on MSPs.  The simple reason for this is Moore's
Law, since computation and the improved digital instrumentation based
on it has dramatically improved our ability to discover and time MSPs.
We need to sample our radio data at rates much faster than the already
rapid spin rates of MSPs in order to make precise measurements of
them.  Effective sampling times of $\sim$50\,$\mu$s are now the
standard for pulsar searches and times below 1\,$\mu$s are common for
high-precision timing observations.

In addition, since radio pulses propagate through the ionized
interstellar medium, they experience dispersion such that the lower
radio frequencies $\nu$ are delayed quadratically (i.e. $\Delta t
\propto \nu^{-2}$) with respect to the high frequencies.  To
compensate, we divide our observing band into many independent
frequency channels, effectively making spectra for each sample in
time, and then delay the channels appropriately so that the pulses sum
in phase.\footnote{For pulsar timing, we can Nyquist-sample the full
  observing band and perform what is known as coherent de-dispersion
  to exactly remove the dispersive delays.  This technique is
  incredibly computationally intensive and has only recently become
  feasible across large observing bandwidths.}  For modern pulsar
surveys, which are (finally) nearly as sensitive to MSPs as they are
to normal pulsars, we require thousands of spectral channels across
our observing bands.  Combined with fast sampling, data rates of
$\sim$50\,MB/s are common for each {\em pixel} in the surveys, and
total data volumes can comprise nearly a Petabyte.  Computing is
extremely important and costly for search processing as well since we
must search over thousands of independent ``Dispersion Measure'' (aka
DM) trials since the amount of dispersion is unknown {\em a priori} for
a new pulsar\footnote{Most modern surveys also perform tens to
  hundreds of so-called ``acceleration'' trials for each DM trial to
  improve sensitivities to interesting pulsars in compact binary
  systems.}.

Finally, since pulsars are intrinsically faint but continuum radio
sources, we want large observing bandwidths to integrate over in order
to maximize the signal-to-noise of our detections.  In general, they
have steep radio spectra which limits the highest useful observing
frequencies to $\sim$3\,GHz, while radio interference, interstellar
scattering, and the Galactic synchrotron background limit low
frequencies to $\sim$300\,MHz.  Custom state-of-the-art digital
instrumentation is required to rapidly sample and channelize
bandwidths of hundreds to thousands of MHz and cluster computing is
required to process it.

The advances of Moore's Law over the last decade have resulted in
pulsar instrumentation which has been asymptotically approaching
perfection given the $\sim$3\,GHz of {\em total} bandwidth available
for pulsar observations.  For the first time ever, we are being
limited by the sizes of our telescopes rather than the capabilities of
the radio receivers or our pulsar instrumentation.  The new
instrumentation has brought new life to ``classic'' single-dish radio
telescopes used for pulsar observations such as Parkes, Jodrell Bank,
and Arecibo, and will finally allow us to use ``new'' telescopes such
as the GBT to their fullest.

Besides dramatically improved sensitivities to new MSPs in the current
generation of pulsar surveys, our ability to time MSPs has increased
in a Moore's Law fashion as well.  In the 30 years since the discovery
of the first MSP by \cite[Backer et al.~(1982)]{bkh82}, the typical
timing precision for high-precision pulsars has improved by a factor
of several hundred, from RMSs of 10s of microseconds to better than
$\sim$100\,ns.  Such timing precision has opened up completely new
probes of physics.

\section{Case in point:  MSP J1614$-$2230}

A beautiful example of how instrumentation can so dramatically change
what is possible in pulsar observing is the MSP J1614$-$2230.  It was
uncovered in a last-generation survey of {\em EGRET} gamma-ray error
boxes \cite[(Crawford et al.~2006)]{crh06} and appeared to be a fairly
``vanilla'' MSP with a spin period of 3.15\,ms in an 8.7-day circular
orbit with a white dwarf.  Because of its position in the sky, it was a
perfect ``test'' pulsar to observe for a minute or two to check the
GBT's observing system before starting regularly scheduled long-term
observations of MSPs in globular clusters near the Galactic center.
After accumulating almost 5 years of not-very-good
(i.e.~$\sim$10\,$\mu$s RMS) timing data in this manner, we noticed
systematic delays during a small portion of the orbit when the pulsar
passed behind the companion star (i.e. superior conjunction).  On
three separate days, the pulses arrived later than expected by
20$-$40\,$\mu$s.  We knew from the orbital parameters and the
Keplerian mass function that the companion star was at least
0.4\,M$_{\odot}$, which is quite massive compared to most other MSP
companions (which are typically 0.1$-$0.2\,M$_{\odot}$).

We conjectured that the systematics were due to the ``Shapiro Delay''
of the pulses as they passed through the gravitational potential of
the white dwarf.  Irwin Shapiro first identified this effect in
1964\cite[]{shapiro64} and then measured it with beautiful radar
experiments in the late 1960's and early 1970's \cite[(e.g.~Shapiro et
al.~1971)]{shapiro71}.  We had just built a brand-new wideband pulsar
instrument for the GBT called
GUPPI\footnote{\url{https://safe.nrao.edu/wiki/bin/view/CICADA/NGNPP}}
(Green Bank Ultimate Pulsar Processing Instrument), based on field
programmable gate arrays (FPGAs) and high-end graphics processing
units (GPUs) made for computer gaming.  GUPPI can effectively
perfectly process the full 800\,MHz bandwidth of the GBT's L-band
receiver, a factor of more than 10 increase in what was possible with
the previous generation high-precision pulsar backend at the GBT.  If
we used it in a week-long observing campaign, with observations of
several hours each day, we predicted that the timing precision would
increase by a factor of up to 10.

The 8-hour observation during conjunction was simply stunning.
Hundreds of data points with $\sim$1\,$\mu$s errors showed the
extremely strong and cusp-like signature indicative of Shapiro delay
from a nearly edge-on orbit.  When the observing campaign was over, we
measured the two parameters associated with the Shapiro delay to high
precision, which in combination with the Keplerian orbital parameters,
gave us the mass of the white dwarf (0.500$\pm$0.006\,M$_{\odot}$),
the orbital inclination (89.17$\pm$0.02\,degrees!), and a pulsar mass
of 1.97$\pm$0.04\,M$_{\odot}$, by far the most massive precisely
measured neutron star to date \cite[(Demorest et al.~2010)]{dpr10}.
The GBT with GUPPI had turned a ``vanilla'' MSP into an important
probe of high-density physics, which has strongly constrained the
neutron star equation of state \cite[(i.e.~EOS; Lattimer \&
Prakash~2010)]{lp10} and touched on many other aspects of both basic
and astro-physics \cite[(e.g.~{\"O}zel et al.~2010)]{ozel10}.  It also
made J1614$-$2230 into a timing array pulsar for the detection of
gravitational waves.

\section{Gravitational waves:  the next frontier for MSPs?}

Pulsar timing arrays (PTAs) have the potential to help revolutionize
our view of the Universe, by giving us direct detections of
gravitational waves (GWs), and maybe (just maybe) doing it before
Advanced LIGO does.  The idea of using pulsar timing to detect
gravitational waves goes back to \cite[Detweiler (1979)]{det79}.  The
basic gist is that GWs with wavelengths of light-years, or
consequently frequencies in the nanohertz regime, will stretch and
compress the space-time through which radio pulses travel and thereby
advance or delay their arrival times here at Earth.  A problem,
though, is that long-term changes in arrival times from a pulsar could
be due to a variety of reasons, such as errors in our atomic
time-standards or planetary ephemerides, or even simply timing noise
from the pulsar itself.  \cite[Hellings \& Downs (1983)]{hd83} though,
showed that the quadrupolar-nature of GWs leads to {\em correlated}
delays in the arrival times from an array of pulsars, based on the
angular separation of pairs of pulsars on the sky.

Where would these GWs come from?  While it is possible that we could
see ``strong'' GWs left over from Inflation or from the interaction of
cosmic strings, most people believe our most likely sources of
nanohertz GWs will come from supermassive (10$^8-$10$^9$\,M$_\odot$)
black hole binaries (SMBHBs) orbiting on years-long timescales before
they coalesce.  Such binary systems are thought to exist throughout
the Universe as a result of galaxy mergers during hierarchical
structure formation \cite[(e.g.~Sesana~2012)]{sesana12}.  These black
hole binaries, even at distances of a Gpc, can cause perturbations of
order 10\,ns in pulsar timing residuals.  If Nature was kind to us,
the strongest of these sources (meaning a combination of the most
massive and closest to us in the nanohertz frequency regime), will be
detected individually.  Such a detection could allow for localization
on the sky, detection in other electro-magnetic wavebands, and might
allow us to ``calibrate'' and improve our PTAs similar to the way that
interferometers calibrate and ``phase-up'' on bright astrophysical
point sources.

Even without a strong nearby individual source, though, we are likely
to detect a steep-spectrum (i.e.~much stronger at lower GW
frequencies) stochastic background made up of the ensemble of all
SMBHBs throughout the Universe \cite[(e.g.~Hellings \& Downs
1983)]{hd83}, within the next 5$-$10\, years.  One of the great things
about the stochastic background is that as we time the pulsars in our
PTAs longer, and therefore to lower frequencies, our sensitivity
increases not just by simply accumulating more data, but by climbing
up the steep spectrum to where the GWs are stronger.  This results in
our sensitivities improving at a much faster rate than the naive
$\sqrt{T}$.  No matter what, the signal amplitudes will only be in the
nanoseconds to tens of nanoseconds regime and so we need the best
instrumentation and the best MSPs for our experiments.

Currently there are three major PTA efforts underway: NANOGrav, the
North American Nanohertz Observatory for Gravitational Waves, which
uses Arecibo and the GBT \cite[(Demorest et al.~2012)]{dem12}; the
European Pulsar Timing Array (EPTA), using several of Europe's largest
telescopes \cite[(van Haasteren et al.~2011)]{epta11}); and the Parkes
Pulsar Timing Array (PPTA), using the Parkes telescope in Australia
\cite[(Manchester et al.~2012)]{man12}.  Each of these efforts has
been timing $\sim$15$-$30 MSPs for the past 5$-$10 years with
continually improving timing residuals, including several pulsars in
the 50$-$100\,ns regime and many more at 100$-$300\,ns.  In addition,
there is an effort underway to combine the data from all three PTAs
into the International Pulsar Timing Array (IPTA; \cite[Hobbs et
al.~2010]{hobbs10}), which will improve the overall sensitivity by a
factor of $\sim$2 compared to any single PTA.

Despite the fact that PTA sensitivities are continuing to get better
due to improvements in our timing techniques and our instrumentation,
we still need more and better MSPs to fully achieve the potential of a
pulsar-based GW ``observatory''.  That potential, and the possibility
of an imminent GW detection by pulsar timing, is driving many
large-scale efforts to find more MSPs.

\section{New surveys for millisecond pulsars}

The millisecond pulsar renaissance can most certainly be seen in the
results of the recent and ongoing surveys for new pulsars.  New
instrumentation and increased compute capacity has dramatically
improved MSP search sensitivities from the same telescopes and even
from the same radio receivers which have surveyed the sky before.  The
number of Galactic MSPs (i.e.~those not in globular clusters, whose
numbers have increased hugely as well) has doubled in the last three
years and quadrupled in the last decade (see Figure~1).  About half of
those found in the past few years have come from a series of
large-area surveys being conducted by the GBT (the GBT Driftscan and
Green Bank North Celestial Cap or GBNCC surveys; Lynch et al.,~these
procs.), Arecibo (Pulsar-ALFA and the AO327 Driftscan surveys; Lazarus
et al.,~these procs.), and Parkes (the three related High Time
Resolution Universe or HTRU surveys; Keith et al.,~these procs.).

\begin{figure}[t]
  \begin{center}
    \includegraphics[trim=30 15 40 25, clip,
        width=0.76\textwidth]{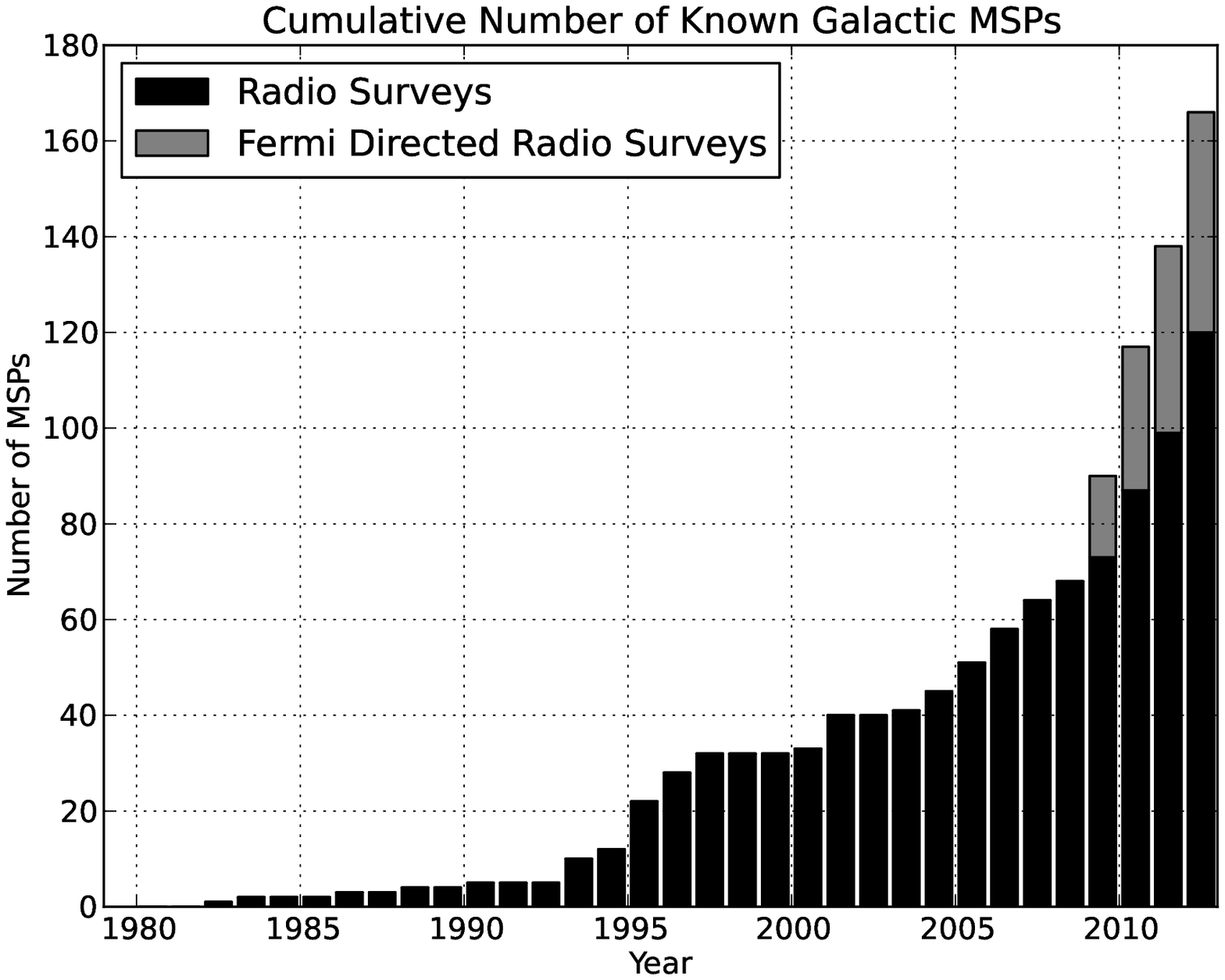}%
    \includegraphics[width=0.23\textwidth]{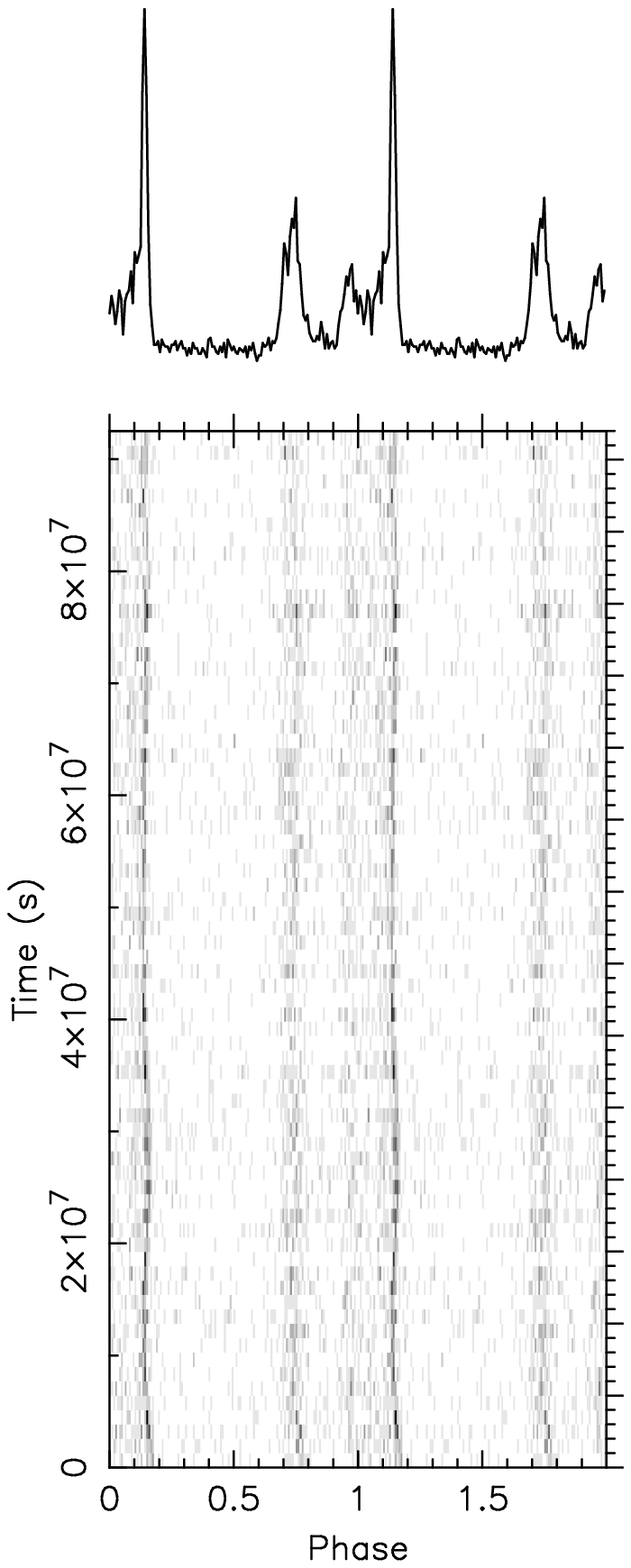}%
    \caption{\textit{(Left)} Number of known Galactic (i.e. not in
      globular clusters) millisecond pulsars (MSPs) as a function of
      year, through November 1, 2012.  MSPs are defined here as
      recycled pulsars spinning faster than 15\,ms.  The rapidly
      increasing numbers of these systems are due primarily to new
      wide-area radio surveys using much improved instrumentation and
      {\em Fermi}'s ability to point us at likely radio MSPs.  For
      up-to-date numbers, see Duncan Lorimer's list at
      \protect\url{http://astro.phys.wvu.edu/GalacticMSPs/GalacticMSPs.txt}
      \textit{(Right)} Folded {\em Fermi} LAT gamma-ray photons of the
      3.68\,ms radio and gamma-ray binary MSP J1231-1411
      \cite[(Ransom et al.~2011)]{ran11}.  The plot contains 3\,yrs of
      {\em Fermi} data, corresponding to $\sim$3000 photons (about 3
      per day!), $\sim$560 binary orbits, and 24 billion rotations of
      the pulsar, yet the main peak has a width of just over 2\% of
      the pulse period --- a beautiful example of the power of pulsar
      timing.}
    \label{fig1}
  \end{center}
\end{figure}

These surveys are using either low-frequency ($\sim$350\,MHz)
receivers with short dwell times or higher-frequency systems
($\sim$1400\,MHz) with multiple beams to increase survey speeds and
enable them to cover large areas of the sky.  When complete in the
next couple years, the full sky will have been re-surveyed by one or
more telescopes and the total data volume will approach 2 Petabytes.
While the data-taking is not quite halfway complete, an even smaller
fraction of the new data has been fully processed, ensuring that the
discovery rate will continue for several years to come.

A fantastic short-cut to finding new MSPs has been provided by the
{\em Fermi} satellite, though.  Shortly after launch, {\em Fermi}
showed that most MSPs are copious producers of pulsed gamma-ray
emission \cite[(Abdo et al.~2009)]{abdo09}.  That meant that many of
the unassociated {\em Fermi} LAT sources, especially those well off of
the Galactic plane, might be MSPs.  A collaboration of radio
astronomers working with the LAT team, called the Pulsar Search
Collaboration, has uncovered at least 45 new radio MSPs (and
counting!) by searching these gamma-ray sources deeply with the
biggest radio telescopes around the world \cite[(Ray et
al.~2012)]{psc12}.  We would eventually find most of these MSPs with
increasingly sensitive all-sky radio surveys, but {\em Fermi} is
showing us exactly where to look and allowing us to find them much
sooner.  Once radio timing solutions are established, basically all of
them are seen to pulse in gamma-rays as well (see Figure~1), giving us
a brand new probe into the pulsar magnetosphere and emission
processes.  A large fraction of the new {\em Fermi} MSPs are turning
out to be previously rare eclipsing systems, a point we currently do
not understand.  Since there seems to be no strong correlation between
the gamma-ray and radio fluxes, we expect that new {\em
  Fermi}-directed radio MSPs will be uncovered for as long as new {\em
  Fermi} sources are being detected.

Most pulsar surveys uncover one or two surprises, and in fact it is
for these exotic systems that we often tune the parameters of our
searches.  High on the current ``Most Wanted'' list are a
sub-millisecond pulsar, which would strongly constrain the EOS of
neutron star matter, and a pulsar-black hole system, which would be an
incredible testbed of strong-field general relativity (see Michael
Kramer's contribution to these proceedings).  We don't have those
yet(?), but there have certainly been other recent surprises.

The GBT Driftscan survey uncovered a fast and bright MSP in a 4.75-hr
orbit which, it turned out, had been studied as a likely accreting
cataclysmic variable, in the optical, radio, and X-rays for the decade
preceding its discovery as an MSP \cite[(Archibald et
al.~2009)]{arc09}.  The pulsar, J1023$+$0038, seems to be a ``Missing
Link'' system in the late stages of the recycling process as stellar
evolution of the evolved companion nears completion.  There are now a
half dozen similar systems known, many uncovered with the help of {\em
  Fermi}, and all fascinating probes into binary and stellar
evolution\footnote{These systems have been coined ``Redbacks'' in
  spider-salute to the ``Black Widow'' pulsars.} (Roberts, these
procs.).

Deep in the Galactic plane, the P-ALFA survey uncovered another
evolutionary oddball, MSP J1903$+$0327, which is in an eccentric orbit
around a Sun-like main-sequence star \cite[(Champion et
al.~2008)]{cha08}.  This highly-inclined system can be timed quite
precisely and has yielded Shapiro delay as well as the relativistic
orbital precession of periastron, thereby providing a very precise,
and fairly high, neutron star mass (1.667$\pm$0.021\,M$_\odot$;
\cite[Freire et al.~2011]{freire11}).  The formation mechanism of the
system is uncertain, but given that the star which spun up the pulsar
is missing, a likely triple scenario involving a dynamical instability
seems to be the best bet.  Intriguingly, the GBT Driftscan survey has
recently uncovered an MSP which is currently in a triple stellar
system (Ransom et al.~in prep.).

Finally, more pulsar planets, or at least planet-mass companions, have
been uncovered in three new MSP systems, two of which were announced
at this conference (see Bates et al.~and Lynch et al.).  The first of
these is the so-called ``Diamond Planet'' system J1719$-$1438, with a
Jupiter-mass companion in a compact 2.2-hr binary \cite[(Bailes et
al.~2011)]{bai11}.  The density of the companion is constrained to be
very large, implying that it is an ultra-low mass carbon white dwarf
in crystalline form (i.e.~diamond!).  Pulsar timing can easily detect
planets of almost any reasonable mass, so it is interesting to ask why
are there so few pulsar planets?

\section{Prospects for the future}

With pulsar instrumentation approaching perfection and several large
surveys underway and already successful, it is easy to predict good
things from pulsars, and in particular MSPs, in the coming few years.
We will eke out additional timing precision from our instruments and
techniques which will lead to more surprises, and potentially on short
timescales --- more planets?  more massive or maybe low-mass neutron
stars?  gravitational waves?  And all-sky or targeted surveys could
uncover a new ``Holy Grail'' any day, an eccentric MSP-MSP binary
could lurk in a globular cluster, and SgrA* should be surrounded by
hundreds of pulsars.

Unfortunately, though, there are some issues.  Nearly all of the
``classic'' pulsar telescopes, Arecibo, Jodrell Bank's Lovell
Telescope, Parkes, and now the GBT, have recently been or are
currently under serious threat of closure due to dwindling or changing
budgets.  ``Simple'' single-dish radio telescopes can do fantastic
things for pulsar astronomy, but they are being eclipsed by current
and next-generation radio arrays which promise to do more things for
more astronomers.  Great pulsar astronomy will certainly be done with
upcoming arrays like MeerKAT, LOFAR, and the Phase I SKA, but things
are trickier and potentially costlier with arrays. It will be
important to carefully weigh the costs of closing simple single-dish
telescopes, which can be very effective at producing high-impact
pulsar science, as we march towards the era of giant radio arrays.

China's upcoming 500-m diameter single-dish called FAST will be an
excellent test case (Li et al.,~these procs.).  Its incredible
sensitivity and increased sky coverage compared to Arecibo could
revolutionize pulsar astronomy before the Phase I SKA is even
partially complete.  No matter what, if and when these giant new
facilities come on line, some of the first and best science you will see
will be from pulsars.  How could it not?  These things are wicked
cool.  Seriously.

\end{document}